# SecMon: End-to-End Quality and Security Monitoring System

Tomasz Ciszkowski[1], Charlott Eliasson[2], Markus Fiedler[2], Zbigniew Kotulski[3], Radu Lupu[4], Wojciech Mazurczyk[1]

[1]Warsaw University of Technology, Institute of Telecommunications, Poland
[2]Blekinge Institute of Technology, Department of Telecommunication Systems, Sweden
[3]Institute of Fundamental Technological Research, Poland
[4]University Politehnica of Bucharest, Romania

**Abstract**

The Voice over Internet Protocol (VoIP) is becoming a more available and popular way of communicating for Internet users. This also applies to Peer-to-Peer (P2P) systems and merging these two have already proven to be successful (e.g. Skype). Even the existing standards of VoIP provide an assurance of security and Quality of Service (QoS), however, these features are usually optional and supported by limited number of implementations. As a result, the lack of mandatory and widely applicable QoS and security guaranties makes the contemporary VoIP systems vulnerable to attacks and network disturbances. In this paper we are facing these issues and propose the SecMon system, which simultaneously provides a lightweight security mechanism and improves quality parameters of the call. SecMon is intended specially for VoIP service over P2P networks and its main advantage is that it provides authentication, data integrity services, adaptive QoS and (D)DoS attack detection. Moreover, the SecMon approach represents a low-bandwidth consumption solution that is transparent to the users and possesses a self-organizing capability. The above-mentioned features are accomplished mainly by utilizing two information hiding techniques: digital audio watermarking and network steganography. These techniques are used to create covert channels that serve as transport channels for lightweight QoS measurement's results. Furthermore, these metrics are aggregated in a reputation system that enables best route path selection in the P2P network. The reputation system helps also to mitigate (D)DoS attacks, maximize performance and increase transmission efficiency in the network.

1. **Introduction**

During the recent years, P2P (Peer-to-Peer) systems have become popular in many domains, such as filesharing (e.g. Napster; Gnutella; eDonkey), content and software distribution (e.g. BitTorrent), and voice and video communication (e.g. Skype, Joost). The latter, in particular, has conquered the communication-via-Internet scene and for many end users Skype is the first choice when using the computer for calling, rather than VoIP based on classical standards (e.g. SIP or H.323). It is expected that P2P systems and related principles will dominate future networks, which makes P2P the given choice, for the system proposed in this paper: SecMon.

As opposed to client-server systems, peers (the nodes of a P2P system) are acting as client, server and potentially even relay at the same time as they form an overlay network. Structured P2P systems distinguish between different types of peers in order to avoid scalability problems. An important feature of P2P systems is self-organisation, which amongst others implies that a peer can enter or leave the network at any time without jeopardising the overall functionality.

For real-time services like VoIP, security and performance are in general antagonists: security mechanisms impose overheads affecting performance, which may make users switch off security measures in order to get better performance [28]. On the other hand, performance monitoring can enable the detection of a security attack, e.g., modification and data tampering by increase of a transmission delay or jitter. In this paper, security mechanisms are used to help with performance and security monitoring.

As mentioned earlier proposed SecMon system is intended for VoIP service deployed in P2P networks. It is characterized by providing security services such as authentication and data integrity based on a lightweight mechanism. Moreover it enables adaptive QoS and (D)DoS attacks detection. It is also a low bandwidth consumption solution, and what is very important is that it is transparent to users while possessing self-organizing capability. These features are achieved by utilizing two information hiding techniques, namely digital audio watermarking and network steganography. These techniques are used to create covert channels that serve as transport channels for lightweight QoS measurement's results.

As mentioned above, this paper considers VoIP service in P2P overlay networks and in this environment the trust management will be used for misbehaviour detection and nodes evaluation. Such evaluation allows creating a long-term node assessment called a reputation. It is a mean that will be used for providing incentives for a good behaviour of nodes and metric, which allows for choosing a more truthful node. In SecMon we also utilize a reputation maintaining system in order to detect any untrustworthy behaviour in the network that interferes with QoS for a VoIP media transmission. In result, in SecMon originators of such activities are isolated. The goal of the reputation management is to motivate nodes to be well cooperative which would improve the network security and performance. In the proposed system the reputation model will be used to improve routing path selection for media transmission as well as for efficient sending of QoS monitoring and security information.

The earlier project AutoMon that has been sponsored by the European Network of Excellence Euro-NGI developed a P2P-based monitoring infrastructure [18], exploiting self-organisation and implementing autonomic networking facilities with focus on generating quality feedback on behalf of the users. Also the earlier authors' research has advocated end-to-end quality monitoring based on passive measurements, e.g. of application-perceived throughput [5, 6, 7, 10, 18]. This monitoring principle requires a continuous exchange of the monitoring results. On the other hand, the use of the suitable summary statistics of packets transmission such as average and standard deviation makes the related monitoring overhead lightweight, especially when it can be piggy-backed onto existing data transfers. The proposed approach enables this in a new way. As pointed out before, the lightweight monitoring targeted by this paper matches well the limited capacity of the covert channels. Moreover, the less capacity needed for performance management, the more is left for providing security: the better the chance for both to coexist.

The paper is organized as follows. In Section 2 we circumscribe the SecMon architecture, node model and a general communication flow between functional blocks. Sections from 3 to 7 describe each defined functional block operations in details. Finally, Section 8 concludes our work.

## 2. SecMon Architecture

In every Peer-to-Peer network one may distinguish two layers: the **Service Layer** which covers functionality of data routing, network discovery, self-organizing logic and a particular service itself, such as VoIP; and the **Transport Layer** as an inherent part of an underlying network, providing data transportation service. SecMon system introduces an intermediate layer which interacts with the Service and Transport layers, as shown in Fig. 1. The detailed description of and Transport and Service layer focuses only on interfaces towards the SecMon system.

SecMon system is designed for delivering lightweight security and QoS guarantees in P2P multimedia networks, in particular VoIP service. It supports the routing process that influences the choice of the best end-to-end paths between two chosen nodes (Source and Destination in Fig. 1), respecting measurements and monitoring of QoS and security along the VoIP call path. The **SecMon Layer** acts as intermediating block and it is implemented on every SecMon-aware peer node.

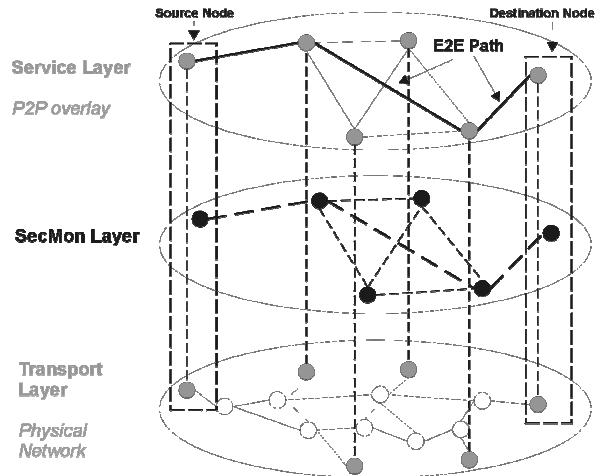

**Fig. 1.** P2P overlay network composition for SecMon

### 2.1 Node Model

Apart from the Data Plane which directly refers to data transmission, all P2P network functionality is organized inside the Control Plane. The SecMon system aims to improve and extend the Control Plane of P2P networks with lightweight security and QoS features.

The P2P network is composed of a certain number of nodes and constitutes a three-layer network model, which is directly reflected in the functional construction of every SecMon node. The node model is depicted in Fig. 2.

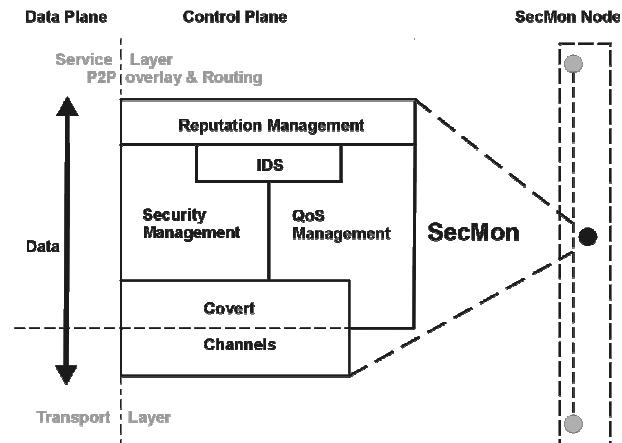

**Fig. 2.** SecMon Layer in a Control Plane of P2P network stack

The SecMon system performs several different operations based on the interaction with outer layers. Inside such a system one may identify internal functions which are specific and organized within the following functional blocks:

- **Covert Channels Block (CCB) –** two sub-channels are used to exchange QoS and Security Measurement data between nodes along the communication path. It is used by SMB and QMB;
- **Security Management Block (SMB) –** is in charge of evaluation of lightweight security measurement data, perform data securing with steganography and watermarking. It delivers metrics to RMB and IDS;
- **QoS management Block (QMB) –** is in charge of measurements and evaluation of lightweight QoS metrics. It delivers metrics to RMB and IDS;

- **Intrusion Detection System Block (IDS) –** anomaly detection system, is obliged to rise alerts in a case of attack detection and report it to RMB;
- **Reputation Management Block (RMB)** – collects and aggregates metrics coming from QMB, SMB, and IDS in order to support a best path-path selection in a P2P routing process.

The SecMon system interactions are realized by means of internal and external interfaces, which have been enumerated in Fig. 3. For an already established VoIP session the SecMon system operations are described by the following messages:

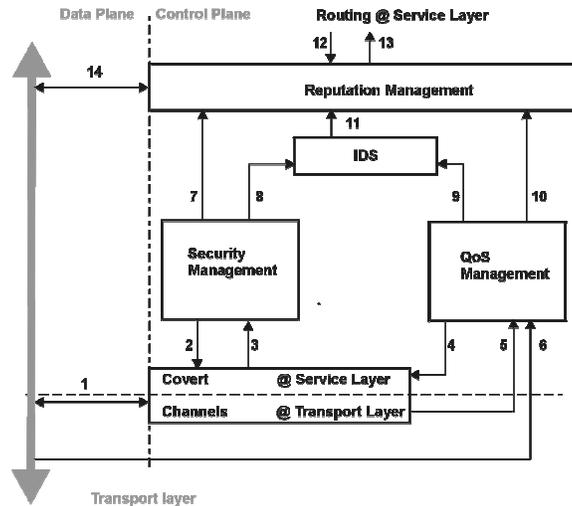

**Fig. 3.** Message flow between SecMon functional blocks

1. Sending/Receiving hidden lightweight QoS and Security measurement throughout data transportation, taking advantage of covert channels and regaining the same bandwidth occupation.
2. Writing lightweight Security measurement data into covert channels.
3. Reading lightweight Security measurement data from covert channels.
4. Writing lightweight QoS measurement data into covert channels.
5. Reading lightweight QoS measurement data from covert channels.
6. Reading measurements parameters (e.g. packet losses, jitter, and throughput) for network performance.
7. Delivering Security metrics to RMB.
8. Delivering Security metrics to IDS.
9. Delivering QoS metrics to IDS.
10. Delivering QoS metrics to RMB.
11. Alerting about (D)DoS attacks to the RMB.
12. Request for the best next hop selection from P2P service layer to RMB
13. Indication answer for the best next hop selection in RMB.
14. Reputation sharing between nodes, with additional communication protocol.

Every SecMon block performs a set of specific operations which are described in details with the following subsections.

### 3. QoS Management Block: Lightweight Monitoring and Related Time Constants

The task of lightweight monitoring consists in discovering performance problems in an efficient way, which means that a rather small amount of monitoring data should be sent and processed in order to derive indications on performance problems. Still, the complete data stream of interest should be observed at different points in the network. These requirements actually exclude active measurements that may impose a significant extra load on links and nodes while merely delivering performance snapshots [7]. Moreover, the proposed method needs to be robust with regards to clock synchronization problems.

In [10] a method for comparative throughput measurements has been proposed, building upon packet or byte counts on short time scales, i.e. for small averaging intervals of length $\Delta T$. From comparisons of the sender's and receiver's throughput histograms, the existence and severity of a bottleneck is seen. Experience has shown that a useful averaging interval should capture at least five to ten packets, while an observation interval $\Delta W$ should capture at least ten averaging intervals. The observation interval can be moved in a jumping or moving window fashion. The method was amongst others used for describing the throughput performance of mobile links [5, 7], and it became obvious that changes of average and standard deviation of the throughput averaged over $\Delta T$ and observed during $\Delta W$ between sender and receiver clearly describe the impact of the bottleneck. This impact can be formulated in utility functions [19], which correlate well with voice Quality of Experience through comparably simple formulae [25, 26]. The statistical calculations can be implemented in a very efficient way using a ring-buffer mechanism [24].

The class of service supported by the SecMon architecture is particularly user-oriented, which makes it necessary to take user-imposed constraints into account. According to a usability analysis [21], glitches of more than 100 ms might be perceived as delays. An average user may loose concentration on a stalling process after about one second. After four seconds, things are getting boring for the user [22]. Users are in general not willing to wait more than ten seconds; otherwise, the risk of abandoning the process is large [21, 22]. The challenge for the SecMon architecture consists in ensuring a reasonably good user perception, which means

- The discovery of performance and security problems preferably within the user reaction time of one second,
- The resolution of such problems within a couple of seconds before the user starts thinking about abandoning the process, e.g. by signalling towards the source so that it can adapt itself to the new conditions, or by finding another route, which we focus on in the sequel.

The combination of its properties and the requirements listed above make the proposed lightweight monitoring method a promising candidate for performance evaluation within SecMon. Its light weight is seen from the following example: Assume (a) $\Delta T = 500$ ms, allowing for the discovery of performance problems on at least that time scale; (b) a sliding window, which means that each $\Delta T$, monitoring data is sent from sender to receiver in order to be compared. Assume a coding of 1 B per piece of data (for instance, average and standard deviation of the packet rate during $\Delta W$); we arrive at a data rate of merely 32 bps.

## 4. Covert Channels Block: Information Hiding Techniques

Information hiding has two sub-disciplines and they are: steganography and digital watermarking. The general difference between these two techniques is that the aim of steganography is to keep the existence of the information secret and in watermarking making it imperceptible.

Steganography is a process of hiding secret data inside other, normally transmitted data, so in ideal situation, anyone scanning data will fail to know it contains covert data. In modern digital steganography, the hidden data is inserted into redundant (provided but often unneeded) data, e.g. fields in communication protocols, graphic image, etc. For TCP/IP steganography (or network stenography) the most important fact is that few fields in the packet's headers are changed during transit.

For digital watermarking the most important application for this project is the ability of embedding the authentication and integrity watermark**.** We can embed data that is similar to a cryptographic hash, into voice samples. This hash will be invisible and inseparable from the data.

### 4.1 Motivation: Secure and Lightweight Transfer of Quality/Security Monitoring Data

By using abovementioned information hiding techniques we will be able to create two covert sub-channels. In this way we may achieve a solution that does not consume transmission bandwidth, because the control bits (that allow distinguishing what data is transferred) are transmitted in a covert (steganographic) channel and data (quality/security monitoring data) is inseparably bound to voice content in a form of digital watermark. Such a usage of covert sub-channels gives opportunity to exchange different types of data (e.g., monitoring information) but it also provides security verification of the transmission source and the content sent (authentication and integrity). Using these covert channels is presented way is described in details in [1] and [2] and will be applied accordingly.

### 4.2 Network Steganography and Its Application

In SecMon for VoIP service we will exploit unused/optional fields in IP/UDP/RTP packets because these protocols are used all IP telephony implementations. Covert data is usually inserted into redundant fields (provided, but often unneeded) for abovementioned protocols and then transferred to the receiving side. As described in [11] and [12] the IP header alone posses a few fields that are available to be used as a covert channel, and we can also deploy unused UDP and RTP protocols fields. The total capacity of those fields exceeds 32 bits per packet.

Furthermore, we can distribute the control bits among these fields in a predetermined fashion (this pattern can be exchanged during a signalling phase of a conversation). In the chosen fields we will only transmit the header (control bits) of our protocol with the use of steganography technique. Because the header consists of only a few bits per packet, such a type of transmission is potentially hard to discover.

### 4.3 Audio Watermarking

The primary application of audio watermarking was to preserve copyright and/or intellectual property protection, sometimes called DRM (Digital Right Management). However, this technique can also be used to create effective a covert channel inside a digital content. Generally, the audio watermarking algorithm consists of two phases: embedding the digital watermark into the voice at the source, and then its extraction at the destination place. In IP telephony we can distinguish those phases too; as soon as the conversation begins, certain information is embedded into the voice samples and sent through the communication channel. Then, the digital watermark is extracted from a voice stream before it reaches a caller.

Currently, there are a number of audio watermarking algorithms available. The most popular methods that can be utilized in real-time communication for VoIP service include: LSB (Least Significant Bit), QIM (Quantization Index Modulation), Echo Hiding, DSSS (Direct Sequence Spread Spectrum), and FHSS (Frequency Hopping Spread Spectrum). For these algorithms the bandwidth of available covert channels depends mainly on the sampling rate and the type of audio material being encoded. In [13], research results have shown that for LSB communication rate is 1 kbps per 1 kHz (e.g. for 8 kHz sampling rate the capacity is 8 kbps), echo hiding around 16 bps, while DSSS achieved 4 bps. Also experiments in [13] have shown that DSSS method's bandwidth is about 22.5 bps, FHSS 20.2 bps, echo hiding 22.3 bps and LSB around 4 kbps. Based on those results one can clearly see that, besides LSB watermarking (which is easy to detect and erase and that is why will not be considered here), the rest of the methods achieved a covert channel bandwidth range from a few to tens of bits per second.

Moreover, most audio watermarking algorithms for the real-time communication are designed to survive the typical non-malicious operations like: low bit rate audio compression, codec changes, DA/AD conversion, or packet loss. For example, in [15], the watermarking scheme developed at the Fraunhofer IPSI (Institut Integrierte Publikations und Informationssysteme) was tested for different compression methods. Results revealed that a large simultaneous capacity and robustness depend on the scale of the codec compression. When the compression rate is high (1:53), the digital watermark is robust only when we embed about 1 bps. With a lower compression rate we can obtain about 30 bps, whereas the highest data rate was 48 bps with good, robust transparent and complexity parameters.

## 5. Security Management Block

In this functional block of the SecMon system, the lightweight security mechanism will operate. This security mechanism will be implemented analogously as defined in [1]. The main concept is based on shared secret scheme – certain tokens (in form of hashes) are calculated locally and then sent to the other communication side, where they are verified. In this way we gain source authentication and data integrity. Moreover, we can also include signalling messages security as a *post-factum* verification method as defined in [1]. Additionally, QoS monitoring data will also be secured, as defined in [2].

## 6. Reputation Management Block

The reputation term is defined as a perceived grade of trustworthiness to a particular peer created by its historical behaviour during observations and interactions with third party peers in the given context and time. This definition describes a concept of peer's reputation expressed by a level of aggregated trust exchanged and shared between other peers. Reputation tends to represent generalized opinions in a local group of peers. As the reputation may comprise an aggregated trust of several network nodes, it becomes a very valuable metric supporting the P2P routing process [4]. In particular the trust has a context, which refers to a specific environment. For SecMon we use the Reputation Management Block to create a trust to the peering nodes, which may influence on the network reliability, expressed by QoS parameters and the Security level.

Presented in Fig. 4, Reputation Management Block is used in SecMon for end-to-end routing support in P2P overlay network. Thanks to the ability of keeping a historical knowledge, the reputations system is eligible to mitigate (D)DoS attacks, maximize performance and increase transmission efficiency in the network.

In SecMon the reputation block operates on sets of information, which are conveyed inside the covert channels. The first is related to QoS metrics of the lightweight monitoring approach and the second refers to security metrics. Based on these the IDS block feeds the RMB with (D)DoS attack reports. Those parameters are monitored by far-end and all intermediates nodes along the communication path, and are used to build an Evidence Repository. Reports are a source of data for Context Reputation Evaluation where Performance and Security Service Reputations are calculated.

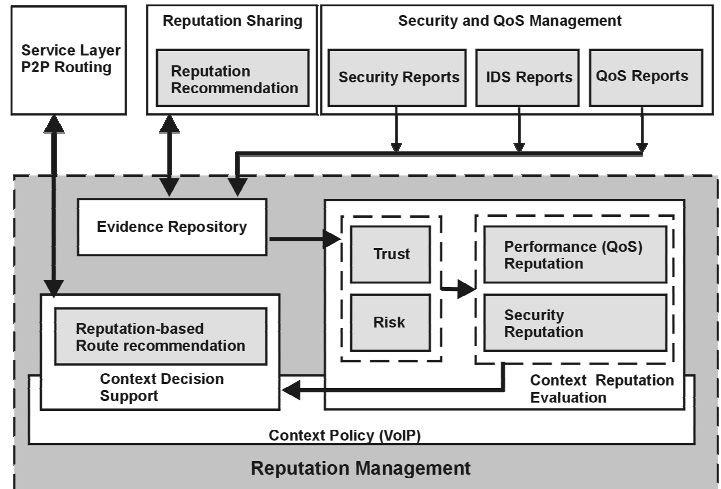

**Fig. 4**. Reputation Management Block for SecMon.

The Performance Service Reputation (PSR) for QoS expresses reliability in the network and is related to the typical quality threats for voice:
- Lost voice samples and packets may imply cracks in and drop-outs of the voice, if no interpolation is effective. Drop-outs longer than one second might entail user reaction. The impact of loss can be seen from changes of the average throughput during $\Delta W$ between sender and receiver (or two arbitrary nodes along the path).
- Delayed voice samples that do not exceed the jitter buffer are uncritical; otherwise, the result is similar to loss. The impact of jitter can be seen from changes of the standard deviation of the throughput with averaging and observation interval $\Delta T$ and $\Delta W$, respectively.
- Too large jitter buffers destroy the feeling of interactivity, but provide the management with some more time to (re-)act before the user actually feels the problems.

The Security Service Reputation (SSR) represents the level of trust for the communicating P2P nodes. Security considerations in the SecMon system are based on the information hiding paradigm. A digital watermarking delivers data integrity and end-to-end authentication. The integrity monitoring and verification may be performed by every node being along the path. This feature allows detecting any malicious activity in the network such as:
- Man-in-the-middle attacks: packet payload modifications, packet retransmission, reply attack, packets reordering, (D)DoS attack.
- Appearing of packet chains in the network, which are not related to the existing sessions, indicating that unauthorized packets in the network consume network resources.

The SSR reflects an opinion on positive and malicious nodes activity. Based on the knowledge coming from evidence repository, the SSR metric is evaluated in order to take preventing actions.

The Reputation-based Route Recommendation function is in charge of indicating the best next hop peer along the path regarding the SSR and PSR values. It interacts with the routing process in the P2P Service Layer supporting VoIP service reliability. Reputation Management Block supports decisions for P2P network resulting in an efficient and best path selection, path re-establishing by selecting alternative routes, misbehaving nodes isolation and service reputation sharing. The reputation may be shared in the local neighbourhood reinforcing the efficient and secure routing.

The reputation service changes in time reacting with possible actions for any detected network issues. In case of alarms triggered by QoS degradation or a security attack, it is reflected by undertaken actions (Table 1).

**Table 1.** Service reputation metrics with a characteristic peer's behaviour and short term reputation system reaction

| PSR - QoS acceptable | SSR – positive hash verification | Conclusion | Short term reputation system reaction |
|---|---|---|---|
| Yes | Yes | Good reputation | Monitoring |
| Yes | No | Security attack | Re-establish new session, isolate affecting nodes |
| No | Yes | Jitter problem | Share reputation with other nodes in order to omit nodes and balance network overloaded nodes |
| No | No | Data losses, network link failure, (D)DoS attack | Re-establish new session |

**6.1 Service Reputation Evaluation**

For SecMon, the reputation evaluation method follows Liu's [17] approach. Based on its adaptation to the peer-to-peer environment [4] the reputation system takes advantage of Own Experience **OE** and second-hand information, which comes from the adjoining peers. The reputation evaluation considers past experience and recommendation reputation of voters (recommends). We define two types of second-hand information, related to the immediate nodes and cumulative reputation describing aggregated reputation of the closest nodes' neighbourhood. The second-hand messages are exchanged on demand of the interested nodes. In order to detect malicious behaviour and any anomalies in the information, second-hand recommendations are validated by a statistical correlation approach.

In SecMon we defined security and performance reputation, SSR and PSR respectively. However, though they reflect two different kinds of activity in the network, the evaluation method is the same for both metrics and corresponds to the Service Reputation **SR.**

Service Reputation evolves in time **n** and for exemplary Fig. 5 is determined by the following equations:

$$SR_n^B(A) = \alpha OE_n^B(A) + (1-\alpha)\frac{\sum_{p \in GVB} IR_n^B(p)V_n^{Bp}(A)}{\sum_{p \in GVB} IR_n^B(p)}$$

$$IR_n^S(p) > 0, \qquad 0 < \alpha < 1, \qquad GVB \equiv peers\ of\ B$$

The Information Reputation **IR** represents credibility of a node recommending other peers. Recommendations **V** come from immediate nodes in the network and reflect the service reputation on other adjacent nodes.

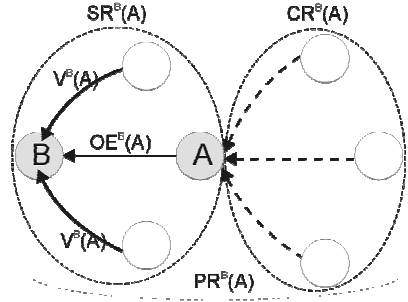
**Fig. 5.** Reputation path building

In order to create the Reputation Path **RP,** which goes throughout nodes B and A, a Cumulative Path **CP** is first calculated:

$$CR_n^B(A) = IR_n^B(A) \frac{\sum_{p \in GA \setminus GB} V_n^{BA}(p)}{|GA \setminus GB|}, IR_n^B(A) > 0$$

$$PR_n^B(A) = SR_n^B(A) CR_n^B(A)$$

## 7. IDS (Intrusion Detection System) Block: Attack Detection Module

Within the SecMon architecture the detection module (IDS) receives input from the QoS and Security Management systems. There are two basic approaches for the attack detection:

- **Signature-based detection:** these techniques assume the existence of a database with all attack definitions (i.e. signatures). The detection mechanism monitors the communications running in the network and compares them against the entries of the signature database. These techniques are efficient when signatures definitions are simple. They usually include packet attributes which can be easily checked. The major drawbacks of these techniques are: the high rate of the false positive alarms, the impossibility to detect flooding attacks based on packets similar to the legitimates ones;

- **Anomaly-based detection:** these techniques assume that in the real world, the malicious attack behaviours progress over the time, which makes the attacks impossible to be foreseen, and consequently detected using the signature techniques. In this regard, these techniques try to build a reference model of the legitimate traffic and raise the alarm anytime the monitored traffic violates this model. Thus, they take the opposite approach from signature techniques. If the model tightly defines the legitimate traffic, it can yield an increased number of false positive alarms. On the other hand, if the model loosely defines the legitimate traffic, it can yield an increased number of false negative alarms. Therefore, one challenge of these techniques is to find out the right set of traffic characterization parameters, in order to trade-off between those two types of false alarms. Another challenge is to keep the legitimate traffic model updated with the daily emerging applications. The main advantage of these techniques is the capacity to detect the newest attacks, even when they are unknown.

The main types of models used for anomaly detection are the following:

- **Behaviour-based model:** which relies on a selected set of traffic parameters whose domain of values is learned through monitoring the attack-free traffic on the network over a predefined period. If one of the parameters' values overcomes a predefined threshold value, the detection system raises an alarm. The sensitivity and accuracy of these models depend on: the quality of the chosen set of traffic parameters, the threshold values, and the granularity of the learning period (e.g. daily, weekly).
- **Standard-based model:** is using the standard specification of communication protocols and the traffic specifications to build an abstract model of the valid traffic. This model can guarantee no false positive alarms, and necessitate a low rate for the model update. It only prevents attacks that violate the standards for communication. The main drawback is the high granularity of monitoring, which needs to be done at the connection level. Consequently, this model requires more resources for processing and storage, for a given level of performance.
- **Misbehaviour detection:** this category meets the characteristics from both categories presented above. These techniques model the malicious behaviour, rather than the good one and operate a detection procedure analogous that for the signature based detection. The main challenge of this category is to find out the proper malicious traffic characteristics which can assure the attack detection from its first stages.

Once an attack has been detected, the SecMon detection module should run the attack characterization task before the reputation system is notified. Attack characterization aims at the definition of the attack's context (w.r.t. offending traffic and the agents of attack). This task yields the input for the SecMon response mechanisms: the reputation system. Therefore, the quality of the attack characterization outcome can influence the performance of the response mechanisms (e.g. collateral losses).

The main properties of the underlying detection system are the accuracy and the timeliness. The accuracy is the property of the system to correctly detect the attacks. The accuracy is measured in number of detection errors achieved by a detection system within a given period. The two categories of detection errors, which we will also take into consideration further on, are false positive and false negative errors.

The timeliness is the property of a detection system to promptly detect the attacks. Since the majority of attacks require some time in order to succeed, a detection window has been defined, within which the detection mechanisms shall work for the early alarm notification. Obviously, the detection window size depends on the type of attack. As soon as the attacks are detected within the detection window, the countermeasures taken by the response system are more effective. When an attack is detected early, the response system can make the attack transparent to the end users, and even avoid the protected network service disruption. Even when the attack is detected late, its negative effects can be reduced and limited.

The overall performance of the detection system depends on its localization. Generally speaking, the accuracy of detection is depreciating proportionally with the distance to the protected network service/resource. This is due the fact that the detection system can only have a complete view of the traffic of the attack close to the protected resource.

Moreover, when detection mechanisms are applied in the network core, they should be less complex, in order to fit the limited resources there. Consequently, the accuracy of detection could be even more diminished. On the other part, when applied close to the potential source of the attack, the detection mechanisms could be quite complex, because the amount of traffic seen at the source of an attack is moderately low, even during the attack.

While placing the detection mechanisms close to the potential target seems to be better solution, the response mechanisms are more effective and incur limited collateral effects when acting close to the source of the attack. However, applying the response mechanisms in the core of the network allows a good trade-off between the costs of the response mechanisms deployment on the high scale and the collateral effects it can produce.

Unfortunately, securing a distributed reactive system proved to be a more complex process and more difficult to achieve, than for an autonomous one (a reactive system whose both mechanisms for detection and response are collocated).

## 8. Conclusion and Future Work

In this paper a lightweight approach for monitoring and managing of QoS and security (SecMon) is introduced and discussed. It is also intended to be used for reputation-based routing in VoIP P2P overlay networks. SecMon implements an additional layer between P2P overlay and (IP-based) transport layer. Moreover, functional blocks of SecMon have been described, which comprise QoS management, security management, IDS, reputation management and covert channels for exchanging management data with other SecMon-enabled peers. It was shown that the latter implements a lightweight solution for QoS management in the sense that no extra capacity needs to be allocated to the monitoring as long as the corresponding data rate remains below some tens of bps, which seems to be feasible. The security management will apply tokens in form of hashes, and the IDS will mitigate (D)DoS attacks of the type flood by the anomaly-based detection built on a behaviour-based model. The core functionality supporting the routing process on the P2P or service layer is the reputation system, taking into account inputs from QoS and security management as well as from the IDS. As the SecMon system offers possibilities to apply attack detection mechanisms (in the form of the reputation-based routing) right in the core of the network, its specific IDS system will allow a good trade-off between the costs of the response mechanisms deployment on the high scale and the collateral effects it can produce.

So far, this paper has introduced a concept of the system which needs to be parameterised, implemented and validated, initially by simulations. For instance, critical timescales in the system need to be linked to the monitoring facilities, while the reputation management block will need to take both long-term, "static" and short-term, "dynamic" [24] performance results into account. The output of the reputation management has to be delivered according to the timescale imposed by the goal for the discovery of performance problems, around one second. Finally, the QoS and security blocks need to act on the level of seconds as described above in order to assure satisfactory end user performance. In this context, the underlying model for the discovery of QoS and IDS problems needs to be specified; one candidate for this is a statistical change point detection algorithm. In the medium term, an implementation involving real users and applications is target in order to demonstrate the viability of the SecMon approach.

**References:**

[1] W. Mazurczyk, Z. Kotulski, *New VoIP traffic security scheme with digital watermarking*, In Proceedings of The 25-th International Conference on Computer Safety, Reliability and Security SafeComp 2006, LNCS 4166, pp. 170 - 181, Springer, Heidelberg 2006.

[2] W. Mazurczyk, Z. Kotulski, *New security and control protocol for VoIP based on steganography and digital watermarking*, Annales UMCS Informatica AI 5 (2006), pp. 417-426.

[3] T. Ciszkowski, Z. Kotulski, *ANAP: Anonymous Authentication Protocol in Mobile Ad hoc Networks*. Proc. 10th Domestic Conference on Applied Cryptography ENIGMA, pp. 191-203. Warsaw, Poland, 2006.

[4] T.Ciszkowski, Z.Kotulski, *Distributed Reputation Management in Collaborative Environment of Anonymous MANETs*, Proceedings of the IEEE International Conference on Computer as a Tool, EUROCON 2007, September 9-12, 2007, Warsaw, Poland, pp. 1028-1033.

[5] S. Chevul, L. Isaksson, M. Fiedler and P. Lindberg, *Measurement of Application-Perceived Throughput of an E2E VPN Connection Using a GPRS Network*, Proc. of $2^{nd}$ EuroNGI IA.8.2 Workshop, Springer (LNCS 3883), Lake Como, Italy, July 2005, pp. 255–268.

[6] S. Chevul. *On Application-Perceived Quality of Service in Wireless Networks*. Licentiate Dissertation, 2006.

[7] M. Fiedler, L. Isaksson, S. Chevul, P. Lindberg and J. Karlsson, *Measurements and Analysis of Application-Perceived Throughput via Mobile Links*, Proc. 3rd Performance Modeling and Evaluation of Heterogeneous Networks (HETNETs) T06, July 18–20, 2005, Ilkley, West Yorkshire, U.K.

[8] J. Mirkovic, S. Dietrich, D. Dittrich, P. Reiher, *Internet Denial of Service: Attack and Defense Mechanisms*, Ed. Prentice Hall, 2004.

[9] *Skype* – http://www.skype.com

[10] M. Fiedler, K. Tutschku, P. Carlsson, and A.A. Nilsson, *Identification of performance degradation in IP networks using throughput statistics*. In: J. Charzinski, R. Lehnert, and P. Tran-Gia (Ed.): Providing Quality of Service in Heterogeneous Environments. Proc. 18th International Teletraffic Congress (ITC-18), Berlin, Germany, Sept. 2003, pp 399–407.

[11] A. Wierzbicki, *Concepts, Methods, and Problems in Trust Management Research*, PJIIT, http://sieci.pjwstk.edu.pl/uTrust.php

[12] S. J. Murdoch, S. Lewis, *Embedding Covert Channels into TCP/IP*. Information Hiding 2005, pp. 247-26.

[13] V. Korjik, G. Morales-Luna, *Information Hiding through Noisy Channels*, Proc. 4th International Information Hiding Workshop, Pittsburgh, PA, USA, (2001) 42-50.

[14] W. Bender, D. Gruhl, N. Morimoto, A. Lu, *Techniques for data hiding*, IBM System Journal, vol. 35, No. 3&4. pp 313-336, 1996.

[15] T. Takahashi, W. Lee, *An Assessment of VoIP Covert Channel Threats*, Proc. 3rd International Conference on Security and Privacy in Communication Networks (SecureComm'07), Nice (France) 17-21 September 2007

[16] M. Steinebach, F. Siebenhaar, C. Neubauer, R. Ackermann, U. Roedig, J. Dittmann, *Intrusion Detection Systems for IP Telephony Networks*, Proc. Real time Intrusion Detection Symposium, Estoril, Portugal (2002) (17)1-9.

[17] J. Liu, V. Issarny, *Enhanced Reputation Mechanism for Mobile AdHoc Networks*, in: Proc. 2nd International Conferenceon Trust Management, iTrust 2004, Oxford, UK, March 29 - April 1, LNCS 2995, pp. 48-62, Springer, Berlin 2004.

[18] Binzenhöfer, K. Tutschku, B. auf dem Graben, M. Fiedler, and P. Arlos, *DNA, a P2P-based framework for distributed network management*. Proc. of 2nd EuroNGI IA.8.2 Workshop, Springer (LNCS 3883), Lake Como, Italy, July 2005.

[19] M. Fiedler, K. Tutschku, S. Chevul, L. Isaksson, and A. Binzenhöfer, *The Throughput Utility Function: Assessing network impact on mobile services*. Proc. of 2nd EuroNGI IA.8.2 Workshop, Springer (LNCS 3883), Lake Como, Italy, July 2005.

[20] S. Chevul, L. Isaksson, M. Fiedler, P. Lindberg, and R. Waltersson, *Network Selection Box: An Implementation of Seamless Communication*. In Proc. of 3rd EuroNGI Workshop IA.8.3 Workshop on Wireless and Mobility, Springer (LNCS 4396), Sitges, Spain, June 2006.

[21] J. Nielsen, *Usability Engineering*. Morgan Kaufman, San Francisco, 1994.

[22] M. Fiedler, ed. *EuroNGI Deliverable D.JRA.6.1.1: State-of-the art with regards to user-perceived Quality of Service and quality feedback*. May 2004. Available at http://www.eurongi.org.

[23] M. Fiedler, S. Chevul, L. Isaksson, P. Lindberg, and J. Karlsson, *Generic communication requirements of ITS-related mobile services as basis for seamless communications*. Proc. of NGI 2005 (published by IEEE), Rome, Italy, April 2005.

[24] L. Isaksson, *Seamless Communications: Seamless Handover Between Wireless and Cellular Networks with Focus on Always Best Connected*. Ph.D. Thesis 2007:06. Blekinge Institute of Technology.

[25] T. Hossfeld, P. Tran-Gia, and M. Fiedler, *Quantification of Quality of Experience for Edge-Based Application*. In Proc. of 20th International Teletraffic Congress (ITC-20), Ottawa, Canada, June 2007.

[26] T. Hoßfeld, A. Binzenhöfer, M. Fiedler and K. Tutschku, *Measurement and Analysis of Skype VoIP Traffic in 3G UMTS Systems*. Proc. of IPS-MoMe 2006, Salzburg, Austria, Feb. 2006, pp. 52–61.

[27] B.E.Brodsky, B.S.Darkhovsky, *Non-parametric methods in change-point problems*, Kluwer Academic Publishers, 1993.

[28] D.R. Kuhn, T.J. Walsh, S. Fries, *Security Considerations for Voice Over IP Systems*, Computer Security Division, Information Technology Laboratory, NIST SP800-58, January 2005.